\documentclass[twocolumn,english,aps,pra,superscriptaddress,showpacs,tightenlines]{revtex4-1}
\usepackage{amsmath}
\usepackage{graphicx}
\usepackage{amssymb}

\begin{document}

\title{Inherent Mach-Zehnder interference with ``which-way'' detection for
single particle scattering in one dimension}
\author{Lan Zhou}
\affiliation{Key Laboratory of Low-Dimensional Quantum Structures and Quantum
Control of Ministry of Education, and Department of Physics, Hunan
Normal University, Changsha 410081, China}
\author{Yue Chang}
\affiliation{Institute of Theoretical Physics, Chinese Academy of Sciences, Beijing,
100190, China}
\author{H. Dong}
\affiliation{Institute of Theoretical Physics, Chinese Academy of Sciences, Beijing,
100190, China}
\author{Le-Man Kuang}
\affiliation{Key Laboratory of Low-Dimensional Quantum Structures and Quantum
Control of Ministry of Education, and Department of Physics, Hunan
Normal University, Changsha 410081, China}
\author{C. P. Sun}
\homepage{http://power.itp.ac.cn/~suncp}
\email{suncp@itp.ac.cn}
\affiliation{Institute of Theoretical Physics, Chinese Academy of Sciences, Beijing,
100190, China}

\begin{abstract}
We study the coherent transport of single photon in a one-dimensional
coupled-resonator-array, ``non-locally'' coupled to a two-level system.
Since its inherent structure is a Mach-Zehnder interferometer, we explain
the destructive interference phenomenon of the transmission spectrums
according to the effect of which-way detection. The quantum realization of
the present model is a nano-electromechanical resonator arrays with two
nearest resonators coupled to a single spin via their attached magnetic
tips. Its classical simulation is a waveguide of coupled defected cavity
array with double couplings to a side defected cavity.
\end{abstract}

\maketitle

\affiliation{Key Laboratory of Low-Dimensional Quantum Structures and Quantum
Control of Ministry of Education, and Department of Physics, Hunan
Normal University, Changsha 410081, China}

\affiliation{Institute of Theoretical Physics, Chinese Academy of Sciences, Beijing,
100190, China}

\affiliation{Institute of Theoretical Physics, Chinese Academy of Sciences, Beijing,
100190, China}

\affiliation{Key Laboratory of Low-Dimensional Quantum Structures and Quantum
Control of Ministry of Education, and Department of Physics, Hunan
Normal University, Changsha 410081, China}

\homepage{http://power.itp.ac.cn/~suncp}

\affiliation{Institute of Theoretical Physics, Chinese Academy of Sciences, Beijing,
100190, China}

\narrowtext

\section{Introduction}

In quantum mechanics, the Bohr's complementarity principle for wave-particle
duality could be displayed in various double-slit experiments (DSE)\cite{www}%
. It manifests that a detection about which way a particle takes in DSE will
inevitably destroy quantum interferences, thus the particle
behavior(spatially localized) emerges while wave-like behavior
(interference) disappears. Otherwise, if the detectors (apparatuses or
environments) can not well distinguish the different possible paths, the
interference recovers\cite{zurek,wwm,ampl}. Lots of experiments have tested
this duality properties through the which-way detection\cite{wwp,wwd,wwa,wws}. In this paper, we show the
effect of which-way detection in a class of experimentally accessible
systems with an inherent structure of the Mach-Zehnder interferometer where
two virtual paths intrinsically are embedded.

\begin{figure}[tbp]
\includegraphics[width=8cm]{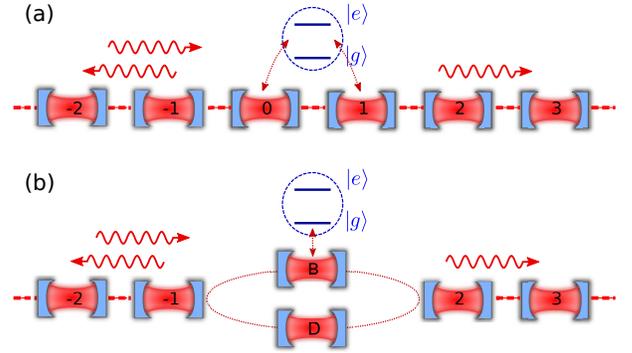}
\caption{(Color online) Schematic of the studied system and the physical
mechanism. (a) The single boson (photon or phonon) scattered by a nonlocal
two-level system simultaneously coupled to two nearest resonators in a
one-dimensional coupled-resonator-array (CRA);(b) the equivalent
Mach-Zehnder interferometer with which-way measurement where the bonding and
anti-bonding superposition of the two local cavity modes form two pathes
for interference, one of which is probed by the TLS while another is
decoupled with it.}
\label{fig1}
\end{figure}

We illustrate this observation with the single particle (photon or phonon)
propagating in a one-dimensional coupled-resonator-array (CRA) where a
localized two-level system (TLS) simultaneously interacts with the confined
EM field modes in two nearest resonators (illustrated in Fig. \ref{fig1}%
(a)). The transported boson is scattered by the TLS when it enters one of
these two nearest resonators, and then the transmission/reflection spectrum
will exhibit the interference pattern of the twice scatterings. One may
superpose these two local EM modes according to their couplings to the TLS
(as illustrated in Fig. \ref{fig1}(b)). Their anti-bonding superposition is
decoupled with the TLS, while the only interaction between the bonding
superposition and TLS makes a which-way detection for the two virtual paths
corresponding to the bonding and anti-bonding modes. In this inherent
Mach-Zehnder interferometer with which-way detector by the TLS, as the
coupling strength increases gradually, the intrinsic interference in the
scattering spectrum (the reflection and transmission spectra) will disappear
progressively. The reason is that the which-way detection only observes the
particle motion in the anti-bonding branch. Therefore the de-interference
phenomenon for a single particle scattering is understood in terms of the
inherent Mach-Zehnder interferometer under the intrinsic {}``which-way''
detection, which seems absent in the real configuration.

The paper is arranged as follows: In Sec. \ref{sec:model},  we introduce the
model for demonstrating the Mach-Zehnder interference. In Sec. \ref{sec:pro},
we study the single photon propagation properties on this structure. The
physical mechanism of the observed phenomena is  explored in Sec. \ref{sec:exp}
by taking advantage of the Mach-Zehnder interference. The possible physical realizations
the system are discussed in Sec. \ref{sec:Physical-implementations}. We conclude the paper in Sec. \ref{sec:concl}.

\section{Model setup \label{sec:model}}

The present system consists of CRA and a side-coupled
TLS. In contrast to the previous model\cite{zhouPRL101}, the TLS
is coupled to two modes supported by two nearest resonators, as illustrated
in Fig. \ref{fig1}(a). The CRA is described by the model Hamiltonian under
the nearest-neighbor approximation
\begin{equation}
H_{C}=\sum_{j}\omega _{c}a_{j}^{\dag }a_{j}-\sum_{j}\xi \left( a_{j+1}^{\dag
}a_{j}+\mathrm{H.c.}\right) ,  \label{ww2-1}
\end{equation}%
with the annihilation/creation operator $a_{j}/a_{j}^{\dag }$ of a confined
boson (photon in a single mode cavity or cooled phonon in a nano-mechanical
resonator) in the jth resonant cavity mode. This is a typical tight-binding boson model.
 For simplicity, we assume that
all cavities have the same eigenfrequency $\omega _{c}$ and the intercavity
coupling has the same strength $\xi $. In this array, waves propagate freely
and are characterized by the dispersion relation
\begin{equation}
\omega _{k}=\omega _{c}-2\xi \cos k,
\end{equation}%
which forms a frequency band in a continuum spectrum.

The TLS with the ground state $\left\vert g\right\rangle $, excited state $%
\left\vert e\right\rangle $ and energy level-spacing $\Omega $, is coupled to two
nearest cavities of the 1D CRA with corresponding strengths $g_{0}$ and $%
g_{1}$. Under the rotating-wave approximation, the dynamics of the TLS
interacting with the electromagnetic field is described by the
Jaynes-Cummings (JC) Hamiltonian%
\begin{equation}
H_{I}=\Omega \sigma _{ee}+\sigma _{-}\left( g_{0}a_{0}^{\dag
}+g_{1}a_{1}^{\dag }\right) +\mathrm{H.c.},
\end{equation}%
where the first term describes the free energy of the TLS with $\sigma
_{ee}=\left\vert e\right\rangle \left\langle e\right\vert $, and $\sigma
_{-}=\left\vert g\right\rangle \left\langle e\right\vert $ and its adjoint $%
\sigma _{+}$ are the corresponding lowing and rising operator. We remark
that the present model can be physically implemented with
nano-electrimechanical resonator array or defect resonator array on photonic
crystal. The detailed discription will be presented in Sec. \ref%
{sec:Physical-implementations}.

\section{Single-photon scattering\label{sec:pro}}

It is clear that the number of excitations is conserved in this hybrid
system. In the one-excitation subspace, two mutually exclusive possibilities
are considered: the particle is either propagating inside the cavity, or
absorbed by the TLS. It indicates that the eigenstate has the form
\begin{equation}
\left\vert E_{k}\right\rangle =\sum_{j}[u_{kj}a_{j}^{\dag }\left\vert
0g\right\rangle +u_{ke}\left\vert 0e\right\rangle ].
\end{equation}%
From the Schr\"{o}dinger equation ($H_{C}+H_{I})\left\vert
E_{k}\right\rangle =E_{k}\left\vert E_{k}\right\rangle $, we derive a series
of coupled stationary equations for the excited-state amplitude $u_{ke}$ and
the amplitudes $u_{kj}$ of single-photon states\cite{zhouPRL101} in the $j-th
$ cavity. Here, the part concerning excited-state
\begin{equation*}
(E_{k}-\Omega )u_{ke}=\left( g_{0}u_{k0}+g_{1}u_{k1}\right)
\end{equation*}%
leads to dispersive coupling strength between the 0th and 1st resonators,
and a non-local effective potential
\begin{equation}
V(j)=G(E_{k})\left( \left\vert g_{0}\right\vert ^{2}\delta _{j0}+\left\vert
g_{1}\right\vert ^{2}\delta _{j1}\right) ,
\end{equation}%
which is proportional to the Green function $G(E_{k})=\left( E_{k}-\Omega
\right) ^{-1}$. The single excitation transport is described by the discrete
scattering equation%
\begin{align}
(\omega _{c}-E_{k})u_{kj}=& \xi \left( u_{kj-1}+u_{kj+1}\right)  \notag \\
+& G(E_{k})\left( g_{j}\delta _{j0}+g_{j}\delta _{j1}\right) \left(
g_{0}u_{k0}+g_{1}u_{k1}\right) .  \label{ww1}
\end{align}%
The first term on the right of Eq.~(\ref{ww1}) charactrizes the hopping between different sites
as the kinetic term. And the double delta potentials in
the send term are induced from the couplings between the TLS and the CRA.
However, such delta potentials are nonlocal comparing with the one-site
coupling in Ref.~\cite{zhouPRL101}. Therefore, the coherent scattering by
the two sites will bring different physical effect.

The process that an incident wave impinges upon the structure, where transmitted
and reflected wave emerge, is formulated by assuming
\begin{equation}
u_{k}\left(j\right)=\exp(ikj)+r_{k}\exp(ikj)
\end{equation}
for $j\leq-1$ and
\begin{equation}
u_{k}\left(j\right)=t_{k}\exp(ikj)
\end{equation}
for $j\geq2$ with reflection and transmission amplitudes, $r_{k}$ and $%
t_{k}$. Concerning Eq.~(\ref{ww1}) to the 0th and 1st resonators with the
above assumptions, we immediately obtain the transmission
\begin{equation}
t_{k}=\frac{2i\sin k\left[\left(E_{k}-\Omega\right)\xi-g_{0}g_{1}\right]}{%
2i\sin
k\left(E_{k}-\Omega\right)\xi-2g_{0}g_{1}e^{ik}-\left(g_{1}^{2}+g_{0}^{2}%
\right)},  \label{ww3-6}
\end{equation}
and reflection amplitudes
\begin{equation}
r_{k}=\frac{2i\sin
kg_{1}^{2}e^{ik}+2g_{0}g_{1}e^{ik}+\left(g_{1}^{2}+g_{0}^{2}\right)}{2i\sin
k\left(E_{k}-\Omega\right)\xi-2g_{0}g_{1}e^{ik}-\left(g_{1}^{2}+g_{0}^{2}%
\right)}.  \label{ww3-7}
\end{equation}
The eigen-energy $E_{k}=\omega_{k}$ is obtained by applying Eq.~(\ref{ww1})
to the resonators far away from the resonators at $j=-1$ to $2$. It is checked that the reflection
coefficient $R=\left\vert r_{k}\right\vert ^{2}$ and transmission
coefficient $T=\left\vert t_{k}\right\vert ^{2}$ satisfy the identity $%
\left\vert r_{k}\right\vert ^{2}+\left\vert t_{k}\right\vert ^{2}=1$.

\begin{figure}[tbp]
\includegraphics[width=7.5cm]{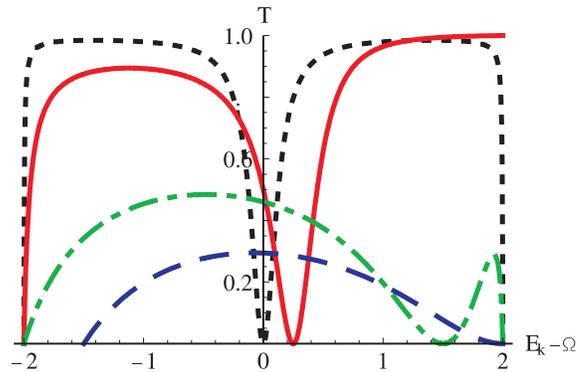}
\caption{(Color online) Transmission coefficient T with transition energy $%
\Omega=\protect\omega_{c}=2$ as a function of the photon incident energy $%
E_{k}$ . The coupling strengths $g_{0}=0.5,g_{1}=0$ for black dotted line, $%
g_{0}=g_{1}=0.5$ for red solid line, $g_{0}=1,g_{1}=1.5$ for green
dot-dashed line, and $g_{0}=1,g_{1}=2.5$ for blue dashed line.}
\label{fig2}
\end{figure}

It is clear that, in Eq.~(\ref{ww3-6}) and~(\ref{ww3-7}), the transmission
generally vanishes at the band edges with $k=0,\pi$. However, constructive
interference is found at $k=\pi$ when $g_{0}=g_{1}=g$ (red solid line in
Fig. \ref{fig2}). Compared with the case of one-site coupling (solid dotted
line in Fig. \ref{fig2}), the transmission does not vanishes at $%
E_{k}=\Omega $ with a nonvanishing amplitude
\begin{equation}
t_{k}=\frac{2ig_{0}g_{1}\sin k}{2g_{0}g_{1}\exp(ik)+%
\left(g_{1}^{2}+g_{0}^{2}\right)},
\end{equation}
due to the {}``non-local'' coupling between the TLS and CRA. Obviously, the
position of transmission zero is shifted to $\Omega+g_{0}g_{1}/\xi$. When
the coupling gets stronger, the transmission dip gradually moves outside the
spectrum band, thus no transmission zero appears, as shown in Fig. \ref{fig2}%
. The coupling between the TLS and CRA induces the effective double delta
potentials in the way of photon propagation. The strength tends to infinite
potentials at $E_{k}=\Omega$, which indicates the total reflection.
However, the straightforward physics picture
fails to explain the nonvanishing transmission at $E_{k}=\Omega$ in the case
of nonlocal coupling.

\section{De-interference and which-way explanation\label{sec:exp}}

To give an intuitionistic explanation about the nonvanishing point of
transmission, we now introduce the bonding and anti-bonding modes
\begin{eqnarray*}
B &=&a_{0}\cos \theta +a_{1}\sin \theta , \\
D &=&-a_{0}\sin \theta +a_{1}\cos \theta ,
\end{eqnarray*}
where $\tan \theta =g_{1}/g_{0}$. When $g_{0}=g_{1}=g$, the modes $B$ and $D$
presents two virtual paths without direct coupling between each other, then
the CRA virtually becomes an equivalent Mach-Zehnder interferometer (see the
Fig.~\ref{fig1}(b)) where the arm $B$ is experienced a which-way measurement
by the TLS. Those cavities on the left of the cavity at $j=-1$ or the right
of the cavity at $j=2$ are regarded as the left and right leads, which are
connected to each other via two arms. The lower arm corresponds to the
anti-bonding model $D$ of eigenfrequency $\omega _{D}=\omega _{c}+\xi \sin
2\theta $, while the upper arm to the bonding mode $B$ of eigenfrequency $%
\omega _{B}=\omega _{c}-\xi \sin 2\theta $ coupling to the TLS with coupling
strength $\sqrt{2}g$. The effective hopping strengthes between the bonding
mode $B$ and nearby cavities are $\xi \cos \theta $ and $\xi \sin \theta $,
as shown in Fig.~\ref{fig1}(b). It is emphasized that anti-bonding mode $D$
is indeed decoupled with the TLS.

In the above equivalent configuration, the propagating state is a
superposition of four mutually exclusive possibilities: i) the particle
propagating inside left and right leads, represent by the state $%
a_{j}^{\dag}\left\vert 0g\right\rangle $ with possibility amplitude $u_{kj}$%
; ii) absorbed by the TLS, by $u_{ke}\left\vert 0e\right\rangle $; iii)
propagating from the left to right lead via the upper arm, by $%
B^{\dag}\left\vert 0g\right\rangle $ with amplitude $u_{kB}$, or iv) via the
lower am, by $D^{\dag}\left\vert 0g\right\rangle $ with amplitude $u_{kD}$.
The superposition of these four possibilities forms a stationary state $%
\left\vert E_{k}\right\rangle $ with the band energy $E_{k}$. Here, three
scattering channels are implied, i.e., single particle travels through the
excited TLS or via the bonding and anti-bonding branches respectively with
the TLS unexcited. In the reduced dimensionality, particles follow two
different paths defined by two virtual modes. We refer to these two paths as
path B and D corresponding to the B-slit and D-slit. A dispersion delta
potential is localized in path B since particle passing the B-slit will
interact with the TLS. The total transmission amplitude is the sum of the
amplitudes in the two branches, and the interference pattern is determined
by the phase difference between the two paths. Therefore, the suppression of
quantum interference depends on the dwell time of the single particle
absorbed by the TLS.

An incoming wave with energy $E_{k}$, incident from the left lead, is split
into two branches at the first junction and joins again into the outgoing
wave at the second junction. The single particle propagating around the ring
(consists of sites $-1,D,B,2$ shown in Fig.~\ref{fig1}(b)) is described as
follows: the discrete scattering equation in the path $B$
\begin{equation}
\left( E_{k}-\omega _{B}\right) u_{kB}=\sqrt{2}gu_{ke}-\frac{\xi }{\sqrt{2}}%
\left( u_{k2}+u_{k-1}\right) ,
\end{equation}%
which is coupled with the local atomic excitation characterized by
\begin{equation*}
\left( E_{k}-\Omega \right) u_{ke}=\sqrt{2}gu_{kB};
\end{equation*}
In the path $D$, the particle propagates freely with motion equation $\left(
E_{k}-\omega _{D}\right) u_{kD}=\xi \left( u_{k2}-u_{k-1}\right) /\sqrt{2}$.
However, on the nodes with$\ j=-1,2$, the amplitudes for the single particle
is coupled to those two splitting nodes $B$ and $D$ with the
following forms respectively
\begin{align}
\left( E_{k}-\omega _{c}\right) u_{k-1}& =-\xi u_{k-2}-\frac{\xi }{\sqrt{2}}%
\left( u_{kB}+u_{kD}\right) , \\
\left( E_{k}-\omega _{c}\right) u_{k2}& =\frac{\xi }{\sqrt{2}}\left(
u_{kD}-u_{kB}\right) -\xi u_{k3}.
\end{align}%
Then, it follows from the above discrete scattering equation that \ the
transmission amplitude $t_{k}$ $=t_{k}^{D}+t_{k}^{B}$ is given by the
partial wave transmission amplitude $\ t_{k}^{D}=i\sin \left( k/2\right)
\exp (-ik/2)$ and
\begin{equation}
t_{k}^{B}=\frac{e^{-ik/2}\xi \left( E_{k}-\Omega \right) \cos \left(
k/2\right) }{\xi \left( E_{k}-\Omega \right) -g^{2}+ig^{2}\tan ^{-1}\left(
k/2\right) }.  \label{ww4-4}
\end{equation}%
respectively in path D (B) in the absence of path B (D).

\begin{figure}[tbp]
\includegraphics[width=7.5cm]{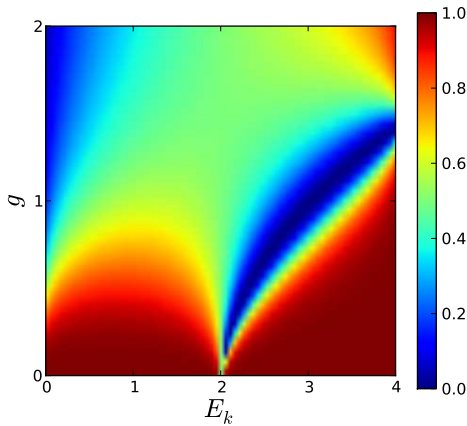}
\caption{(Color online) Transmission coefficient $|t_k|^2$. Its contour plot with
respect to $E_{k}$ and coupling strength $g$. All parameters are in units of
$\protect\xi$. }
\label{fig3}
\end{figure}

When the incident single particle is resonant to the TLS, the spontaneous
emission from the TLS and the propagating modes in the 1D continuum lead to
the complete suppression of the wave transmission in path B, then photons
take path D. Therefore, the TLS prevents single-particle interference of
paths B and D. In this sense, the TLS serves as a which-path detector. It is
the potential exerted by the TLS that makes waves accumulate a phase on path
B, then the wave interference of paths B and D displays a transmission zero
in $t_{k}$. However, the coupling strength characterizes the time that the
single quantum dwells in the TLS. Consequently, the interference pattern is
expected to diminish as the coupling strength increases. Figure \ref{fig3}
shows the contour plots of transmission coefficient $|t_{k}|^{2}$ as a
function of the incident energy $E_{k}$ and coupling strength $g$. Here, we
can see that as the coupling strength $g$ increases the complete suppression
of the wave transmission begins at $E_{k}=\Omega$, gradually shifts to the
band edge, and then disappears. One may also find that the perfect
transmission at $k=\pi$ is independent of the potential exerted by the TLS.
When $k=\pi$, the probability amplitudes $u_{kj}=(-1)^{j}$ for $j\neq0$, $1$%
, lead to destructive interference at B-slit since $u_{kB}=0$, and
constructive interference at D-slit. In this case, the TLS is effectively
decoupled from the CRA.

\section{Physical implementations\label{sec:Physical-implementations}}

\begin{figure}[tbp]
\includegraphics[clip,width=7cm]{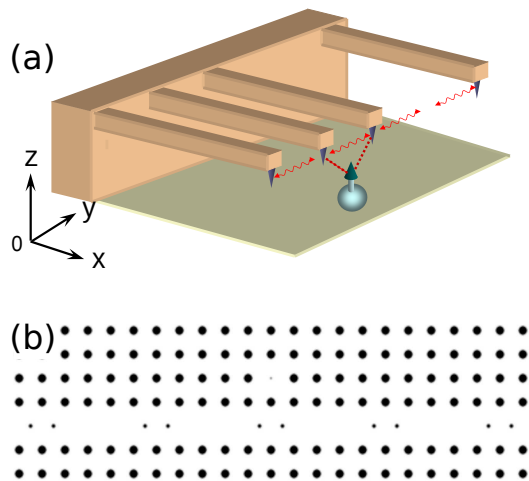}
\caption{(Color online)Two physical implementations: (a).
Nano-electromechanical resonator arrays where two nearest resonators with
ferro-magnetic particles in the tips are coupled to a localized spin. The
origin of the coordinate frame is at the spin. Here, all the resonators are
charged so that they interact with neighbor ones via Coulomb forces; (b). In
the 2D photonic crystal, a side-defected cavity with double couplings to a
waveguide of coupled defected cavity arrays. In the single excitation
subspace, the side cavity with two states behaves as a TLS.}
\label{fig4}
\end{figure}

The above {}``non-local'' coupling obviously distinguishes the present
investigation from previous extensive ones~\cite%
{FanPRL95,zhouPRL101,zhouPRA74,necs2010} where the TLS only couples with the
EM field in a single cavity. Experimentally, the previous setups are
feasible implemented with the confined photon or the single mode phonon in
some confined nano-structures~\cite{pcmfl}, e.g., the circuit QED system, a
semiconductor quantum dot coupling with nanoscale surface plasmons or the
defect cavities of photonic crystal, and the nano-electromechanical
resonator arrays where every resonator is coupled to a localized spin~\cite%
{nanotip}. To implement the present {}``non-local'' coupling seems
difficult in the photonic CRA, but it could be feasible for the
nano-electromechanical resonator array coupled to a local spin, where two of
the resonators are attached by magnetic tips producing magnetic fields in
the x-direction (see the Fig. \ref{fig4}). With the charged resonators
oscillating in the z-direction, and under the rotating-wave approximation,
the inter-resonator coupling is realized via Coulomb forces~\cite{nanotip},
while two of them simultaneously interacts with a single spin through the
magnetic-field gradients \cite{Changy}. In nano-electromechanical resonator
experiments, the parameters $\omega_{c}$, $\xi$ and $\Omega$ can easily
reach a million Hertz frequency scale, and the cavity-TLS coupling strength
is in the order of 100kHz. The second implementation of the {}``non-local
coupling'' could be in a 2D photonic crystal~\cite{dzxu}, which is made up
of a square lattice of high-index dielectric rods, as illustrated in Fig. ~\ref{fig4}. Here, two nearest
cavities in a waveguide of coupled defected cavity arrays is coupled to a
side defected cavity. In the single excitation subspace, the side cavity
with two states (vacuum and single photon) simulates the TLS in our general
setup.

\section{Conclusion \label{sec:concl}}

In conclusion, we have studied the effect of which-way detection inherent to
a class of experimentally accessible systems with some intrinsic
Mach-Zehnder interferometer configuration to enjoy the quantum/classical
interference of two virtual paths. This observation is used to explain the
discovered progressive de-interference in the transmission spectrum of
single photons propagating in 1D CRA, as the non-local couplings of a TLS to
two nearest resonators increases its strength gradually. Besides the quantum
realizations with the nano-electromechanical resonator arrays where two
nearest resonators with magnetic tips simultaneously interact with a single
spin, the classical analogue is proposed based on a waveguide of coupled
defected cavity array with double couplings to a side defected cavity.

This work was supported by NSFC through grants 10974209, 10935010, 11074071,
and 10775048, the National 973 program (Grant No. 2012CB922103). NCET-08-0682, 
PCSIRT No.~IRT0964, the Project-sponsored by SRF for ROCS, SEM {[}2010{]}609-5, 
and the Key Project of Chinese Ministry of Education (No. 210150).


\begin{thebibliography}{99}
\bibitem{www} N. Bohr, in Albert Einstein: Philosopher Scientist (ed.
Schilpp, P. A.) 200 (Library of Living Philosophers, Evanston,1949).

\bibitem{zurek} W.H. Zurekk, Rev. Mod. Phys. \textbf{75}, 715 (2003).

\bibitem{wwm} A. Stern, Y. Aharonov, and Y. Imry, Phys. Rev. A \textbf{41},
3436 (1990).

\bibitem{ampl} Y. Aharonov, D.Z. Albert, and L. Vaidman, Phys. Rev. Lett.
\textbf{60}, 1351 (1988).

\bibitem{wws} E. Buks \textit{et al.}, Nature \textbf{391}, 871 (1998)

\bibitem{wwa} S. Durt, T. Nonn, and G.Rampe, Nature \textbf{395}, 33 (1998).

\bibitem{wwd} R. Schuster \textit{et al.}, Nature \textbf{385}, 417 (1997) .

\bibitem{wwp} X.Y. Zou, L. J.Wang, and L. Mandel, Phys. Rev. Lett. \textbf{67%
}, 318(1991).

\bibitem{FanPRL95} J.T. Shen and S. Fan, Phys. Rev. Lett. \textbf{95},
213001 (2005).

\bibitem{zhouPRL101} L. Zhou \textit{et al.}, Phys. Rev. Lett. \textbf{101},
100501 (2008).

\bibitem{zhouPRA74} L. Zhou,F.M. Hu, J. Lu, and C.P. Sun, Phys. Rev. A
\textbf{74}, 032102 (2006); L. Zhou, Y.B. Gao, Z. Song, and C.P. Sun, Phys.
Rev. A \textbf{77}, 013831 (2008). L. Zhou, H. Dong, Y.X. Liu, C.P. Sun, and
F. Nori, Phys. Rev. A \textbf{78}, 063827 (2008). L. Zhou, S. Yang, Y.X.
Liu, C.P. Sun, and F. Nori, Phys. Rev. A \textbf{80}, 062109 (2009); J.Q.
Liao et al, Phys. Rev. A \textbf{81}, 042304 (2010).

\bibitem{necs2010} O. Astafiev \textit{et al.}, Science \textbf{327 }, 840
(2010).

\bibitem{pcmfl} J.D. Joannopoulos \textit{et al.}, \textit{Photonic crystals
molding the flow of light}, Princeton university press (2008).

\bibitem{nanotip} P. Rabl \textit{et al.}, Nature Phys. \textbf{6}, 602
(2010).

\bibitem{Changy} Y. Chang, C.P. Sun, e-print arXiv:1101.2115.

\bibitem{dzxu} D.Z. Xu, H. Ian, T. Shi, H. Dong and C.P. Sun, Sci. China
Phys. Mechanics and Astronomy \textbf{53}, 1234 (2010)
\end{thebibliography}
\end{document}